\documentclass[twocolumn,authoryear]{aastex63}
\usepackage{appendix}

\usepackage{graphicx}	% Including figure files
\usepackage{amsmath}	% Advanced maths commands
\usepackage{amssymb}	% Extra maths symbols
\usepackage{natbib}
\usepackage{times}
\usepackage{multirow}
\usepackage{subfigure}
\usepackage{url}
\usepackage{etoolbox}

\shortauthors{Maiti et al.}

\begin{document}
\title{Cosmic ray transport in Magnetohydrodynamic turbulence}
\correspondingauthor{Huirong Yan, huirong.yan@desy.de}

\author[0000-0002-0786-7307]{Snehanshu Maiti}
\affiliation{Deutsches Elektronen-Synchrotron DESY \\
Platanenalle 6, D-15738 \\
Zeuthen, Germany}
\affiliation{Institute fur Physik und Astronomie \\
Universitat Potsdam, Haus 28, Karl-Liebknecht-Str \\
24/25,D-14476 Potsdam, Germany}

\author{Kirit Makwana}
\affiliation{Department of Physics,  \\
Indian Institute of Technology Hyderabad \\
 Medak, Telangana 502285, India}

\author[0000-0003-2840-6152]{Heshou Zhang}
\affiliation{Institute fur Physik und Astronomie \\
Universitat Potsdam, Haus 28, Karl-Liebknecht-Str \\
24/25,D-14476 Potsdam, Germany}
\affiliation{Deutsches Elektronen-Synchrotron DESY \\
Platanenalle 6, D-15738 \\
Zeuthen, Germany}

\author[0000-0003-2560-8066]{Huirong Yan}
\affiliation{Deutsches Elektronen-Synchrotron DESY \\
Platanenalle 6, D-15738 \\
Zeuthen, Germany}
\affiliation{Institute fur Physik und Astronomie \\
Universitat Potsdam, Haus 28, Karl-Liebknecht-Str \\
24/25,D-14476 Potsdam, Germany}

\begin{abstract}
%\nolinenumbers
This paper studies cosmic ray (CR) transport in magneto hydrodynamic (MHD) turbulence. CR transport is strongly dependent on the properties of the magnetic turbulence. We perform test particle simulations to study the interactions of CR with both total MHD turbulence and decomposed MHD modes. The spatial diffusion coefficients and the pitch angle scattering diffusion coefficients are calculated from the test particle trajectories in turbulence. Our results confirm that the fast modes dominate the CR propagation, whereas Alfv\'en and slow modes are much less efficient and have shown similar pitch angle scattering rates. We investigate the cross field transport on large and small scales. On large/global scales, normal diffusion is observed and the diffusion coefficient is suppressed by $M_A^\zeta$ compared to the parallel diffusion coefficients, with $\zeta$ closer to 4 in Alfv\'en modes than that in total turbulence as theoretically expected. For the CR transport on scales smaller than the turbulence injection scale, both the local and global magnetic reference frames are adopted. Super-diffusion is observed on such small scales in all the cases. Particularly, CR transport in Alfv\'en modes show clear Richardson diffusion in the {\em local} reference frame.  The diffusion transition smoothly from the Richardson's one with index 1.5 to normal diffusion as particles' mean free path decreases from $\lambda_\|\gg L$ to $\lambda_\|\ll L$, where L is the injection/coherence length of turbulence. Our results have broad applications to CRs in various astrophysical environments.
\end{abstract}


\section{Introduction}

Magnetohydrodynamic (MHD) turbulence is ubiquitous in astrophysical plasmas ranging from interplanetary space to interstellar and intergalactic medium. Propagation of cosmic Rays (CRs) are determined by their interactions with the magnetic turbulence. 
Unlike hydrodynamic turbulence, MHD turbulence can be decomposed into three plasma modes: Alfv\'enic modes, and magneto-sonic slow and fast modes \citep{CL02_PRL}. The Alfv\'en and slow modes have shown scale-dependent anisotropy \citep{GS95, LG01}, whereas fast modes are much more isotropic \citep{CL03, KY20}. The scattering of CRs can be characterized by their interaction with the three MHD modes \citep[see, e.g.,][]{Schlickeiser02}. Because of anisotropy of Alfv\'enic turbulence, the cosmic ray scattering and acceleration in turbulence has been demonstrated to be dominated by compressible MHD modes, particularly the isotropic fast modes \citep{YL02,YL04,YL08,YLP08}. CR transport, therefore, depends much on the MHD modes composition of turbulence, which varies depending mostly on the driving mechanism of MHD turbulence \citep{KY20}. 

%, the damping, which is determined by plasma beta, Mach number as well as collisionality of the medium
%This is important revision to the early assumption that CRs are interacting with hydrodynamic turbulence and transport homogeneously. Propagation of CRs could also be described in terms of CR energy, initial pitch angle and its mean free path in turbulence. Particle diffusion and scattering are studied via the spatial diffusion coefficients and pitch angle scattering coefficients. 

The cross field transport is also shown to differ in the tested model of turbulence than from earlier scenario. A popular concept before was that CR undergoes subdiffusion owing to field line random walk (see e.g., \citealt{KJ2000}). Observations of solar energetic particles in the heliosphere indicated a faster diffusion process perpendicular to the solar magnetic field. The magnetic field separation in Alfv\'enic turbulence presents a close analog of the separation of particles in turbulent media due to the well known process of Richardson diffusion \citep{Richardson26, LVC04,MCB04}). It is demonstrated that subdiffusion does not apply and instead CR cross field transport is diffusive on large scales and superdiffusive on small scales \citep{YL08}. Superdiffusion has also been observed in both the solar wind \citep{Perri2009} and the supernova remnant in interstellar medium \citep{Perri2016}. 

Because of nonlinear nature of turbulence, it is necessary to test the theories with numerical simulations. Earlier numerical simulations  employed only total MHD turbulence and the global magnetic field as the reference \citep{BYL2011,XY13}. We conduct systematic studies of particle transport in this paper, comparing the contributions from total MHD turbulence as well as those from individual MHD modes. The latter is crucial since  proportion can vary in real astrophysical environment \citep{Zhang20NA} and this can be a key factor in determining the local diffusion coefficient, which can substantially deviate from the Galactic mean value \citep[see, e.g.,][]{HAWC_Geminga}. We also adopt reference frames defined by both the global and local magnetic magnetic fields since they generally differ in turbulence environment. The CR transport in the local reference frame is essential for study of particle transport on small scales, e.g., near sources where CRs are freshly injected \citep[e.g.][]{LYZ19PRL}. Also different from earlier studies, we focus on particles of Lamor radii within the inertial range so that the test particle simulation results can be directly confronted with the theoretical results. 

In this paper, we perform test particle simulations in Section 2. We investigate how pitch angle scattering differs among different modes in Section 3. We present the results of perpendicular and parallel diffusion of CRs on global scales in Section 4. The perpendicular transport of CRs on scales smaller than the turbulence injection scale is studied in Section 5. Our results are summarized in Section 6.

\section{Numerical setups}
We have performed 3D MHD simulations to generate turbulence data cubes using two types of MHD codes: the one based on \citet{CL02_PRL}; the other with PENCIL codes \footnote{Please see \url{http://pencil-code.nordita.org} for details.}. The turbulence data cubes are set with $L_{box}^3=512^3$ resolution, energy injection scale $L\sim 0.4L_{box}$. The 3D turbulences are driven by solenoidal forcing. Upon the full development of MHD turbulence, snapshots of turbulence are employed in the test particle simulations. We modulate the external mean magnetic field to produce MHD turbulence with different Alfv\'enic Mach number $M_A$, defined by:
\begin{equation}
M_A\equiv<\frac{\delta V}{V_A}>\sim\frac{\delta B}{B_0}
\end{equation}
Here the quantity $V_A$ is the Alfv\'enic velocity, the symbol ``$<...>$'' indicates the spatial average, $\delta B$ and $B_0$ are turbulent and average magnetic field, respectively.
The $M_A$ values of the generated turbulence data cubes are listed in Table~\ref{Tab:Machnumber}. We also consider the CR transport in decomposed MHD modes. Based on the method described in \citet{CL02_PRL,CL03}, the 3D MHD turbulence is decomposed into three eigen modes,  Alfv\'en, slow and fast. The energy of the magnetic fluctuations for the decomposed modes are normalized to the same amplitude as the total turbulence data cubes. Hence, the decomposed modes and the total turbulence have the same $M_A$.

\begin{table}[ht]
\centering
\begin{tabular}{p{0.7cm}p{0.7cm}p{0.7cm}p{0.7cm}p{0.7cm}p{0.7cm}p{0.7cm} }
\hline
\hline
\multicolumn{7}{c}{$M_A$ for simulations in whole turbulence data} \\
\hline
 0.44 & 0.56 & 0.65 & 0.83 & 1.28 &1.40 &1.54   \\
\hline
\multicolumn{7}{c}{$M_A$ for tests in decomposed turbulence data} \\
\hline
 0.40 & 0.50 & 0.65 & 0.68 & 0.73  & 0.80 & 0.91    \\
\hline
\end{tabular}
\caption{This table lists the Alfv\'enic Mach numbers of turbulence data cubes employed in our simulations. The first row are data cubes only for the test particle simulations with whole turbulence data cubes. The second row are the data cubes with decomposition performed.}\label{Tab:Machnumber}
\end{table}

Test particle simulations are carried out in the MHD turbulence to trace the trajectories of CRs. Since the relativistic particles considered in our study have speed much larger than the Alfv\'en speed, the magnetic fields are considered as stationary and the electric fields in the turbulent plasma are neglected for the study of CR transport. The particle motion is governed by:
\begin{equation}
m\gamma  \frac{d{\bf v}}{dt} = q({\bf v\times B})
\end{equation}
where q, m and {\bf v} represents the charge, mass and velocity of the particles, respectively. $\gamma$ is the Lorentz factor and {\bf B} is the turbulent magnetic field. The Larmor radius of the particle is $r_L=mc^2\gamma/eB_0$. In the current simulations, the dissipation scale of turbulence is $\lesssim0.02$ cube size. We choose $r_L=0.04$ so that it well resides within the inertial range. The particle trajectory tracer follows the Bulrisch Stoer method \citep{PFT86}. The periodic boundary conditions are adopted. The CR diffusion coefficients are calculated from the particle trajectory data. 

\begin{figure}[htp]
\includegraphics[width=\columnwidth]{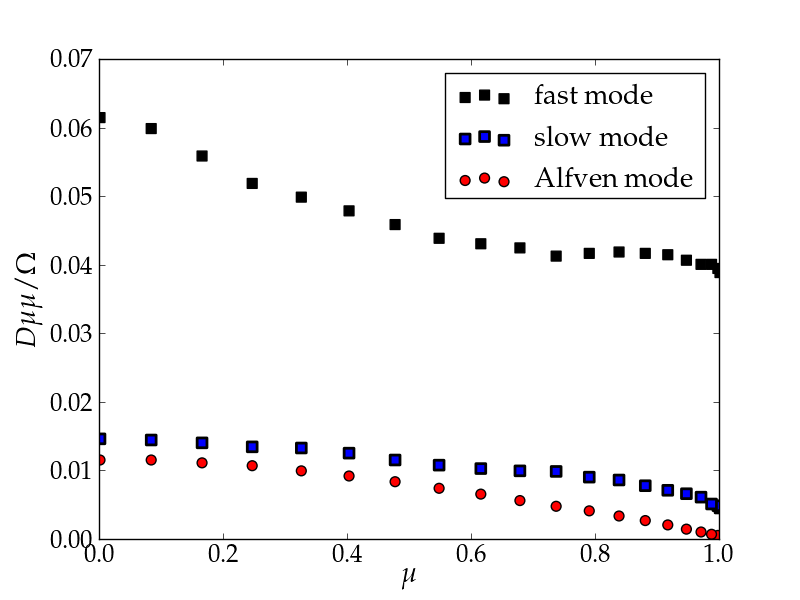}
\centering
\caption{ Pitch angle diffusion coefficients for CRs in different MHD modes with $M_A\sim 0.9$. The x axis represents the initial pitch angle cosine, $\mu$. The y axis represents the pitch angle scattering coefficient normalised by the gyrofrequency, $D_{\mu\mu}/\Omega$. Different symbols represent different MHD modes: Alfv\'en (red), slow (blue) and fast (green).}
\centering
\label{scatt_allmodes}
\end{figure}

\section{Pitch angle scattering}

CR pitch angle scattering is studied here by injecting CRs into MHD modes with the same initial pitch angle $\mu_0$ but random initial positions. The initial pitch angle cosine $\mu_0$ is varied from 0 to 1 with an interval of $0.05$. 10000 particles are used for each test run. The simulations are performed for a few particle gyro periods so that the pitch angle deviation is small (i.e., the root mean square deviation of pitch angle cosine is between 0.01 to 0.1). The pitch angle diffusion coefficient is defined by
\begin{equation}
D_{\mu\mu}=<\frac{(\mu-\mu_0)^2}{2t}>
\label{eq:dmm}
\end{equation}
where the results is averaged among the particles.

The pitch angle diffusion coefficients and their variation with initial pitch angle cosines is presented in Fig.\ref{scatt_allmodes} for the three MHD modes, Alfv\'en, slow and fast.
The result agrees well with the prediction of the nonlinear theory in YL08 \citep{YL08}. Compressible modes contribute to particle scattering through both gyroresonance and resonant mirror (transit time damping, TTD) interaction, the latter of which only operates with compressible modes. Alfv\'en modes, on the other and, only scatter particles through gyroresonance. This is why slow modes are slightly more efficient in scattering particles despite that they have the similar anisotropy as Alfv\'en modes. In comparison to the anisotropic Alfv\'en and slow modes, the scattering with the isotropic fast modes are more efficient. We note that the inertial range in the current MHD simulations is limited. The interstellar turbulence cascade spans more than 10 decades \citep{Armstrong95, Chep2010}. CRs experience, therefore, much more anisotropic Alfv\'enic turbulence on the resonant scales, which are 6-7 orders of magnitude smaller than the turbulence injection scale ($\sim 100pc$) in interstellar medium. This indicates the role of fast modes in scattering CRs is even more prominent in the Galactic ISM.

The pitch-angle scattering determines the diffusion of CRs parallel to the magnetic field. By inserting $D_{\mu\mu}$ from our simulation into the following equation, the parallel mean free path of CR ($\lambda_\|$) can be calculated \citep{Earl74}:
\begin{equation}\label{mfp1}
\frac{\lambda_\|}{L} = \frac{3}{4} \int_{0}^{1}\frac{d\mu v (1 - {\mu}^2 )^2} {D_{\mu\mu} L}
\end{equation}
This calculation will be further cross-checked with the parallel diffusion discussed in the next section.

\section{Particle transport on large scales}

The CR diffusion is strongly dependent on the transportation scale, i.e., larger or smaller than the magnetic coherence length of the turbulence (at the injection scale $L$ for our simulations). Hence, we will separate the calculations for the two cases and only focus on the large scale transport in this section.

We set random initial position and random initial pitch angles for CRs with large scale transport. 2000 particles are used for each simulation. The simulations are run in total turbulence and decomposed MHD modes with different Alfv\'enic Mach numbers. The simulations are carried out for thousands of particle gyro periods until a normal diffusion regime is reached (see Fig.~\ref{fig:dxx} in Appendix for details).

In the current simulations the mean magnetic field is in the x direction, hence the perpendicular diffusion coefficient ($D_{\perp}$) is calculated as:
\begin{equation}\label{eq:dperp}
D_{\perp} =< \frac{(y - {y_0})^2 +(z - {z_0})^2}{2t} >
\end{equation}

The parallel diffusion coefficient ($D_{\parallel}$) is calculated as:
\begin{equation}\label{eq:dpara}
D_{\parallel} =< \frac{(x - {x_0})^2}{2t} >
\end{equation}

The parallel mean free path of particles $\lambda_\|$  is related to $D_{\parallel}$ by $\lambda_\|=3D_{\parallel}/v$. 
The $\lambda_\|$ calculated this way is comparable to those obtained from $D_{\mu\mu}$ in Eq.~\ref{mfp1} and presented in Table~\ref{Tab:mfp_comp}.

\begin{table}[ht]
\begin{center}
\begin{tabular}{ cccccccc  }
\hline
\hline
$M_A$ & 0.4 & 0.5 & 0.65 & 0.68 & 0.73 &0.8 &0.91 \\
\hline
  $\lambda_\parallel$ from $D_\parallel$   & 12.7    & 10.25   &   9.3 & 5.8  & 3.5 & 1.92  & 1.2 \\
 $\lambda_\parallel $ from $D_{\mu\mu}$      & 12.5    & 10.03   &   9.1 & 5.9  & 3.3 & 1.9 & 1.15 \\
\hline
\end{tabular} 
\end{center}
\caption{ 
The mean free path $\lambda_\parallel$ (in unit of box size) is calculated from $D_\parallel$ and $D_{\mu\mu}$ and is listed in the table. The mean free paths in the unit of box size obtained are similar in value from both the above methods of calculation.}\label{Tab:mfp_comp}
\end{table}

The mean free path in the simulations is large, $\lambda_\| > L$, due to the limited numerical resolution and therefore limited inertial range. The regime $\lambda_{\|} > L$ corresponds to the transport of ultra high energy CRs and high energy Galactic CRs in molecular clouds.

On the other hand, the mean free path for most Galactic CRs is smaller than the energy injection scale ( $\lambda_\| < L$). In order to study this regime, more scatterings are artificially introduced in the test particle simulations. In each time step of particle motion, the pitch angle scattering is artificially boosted by a constant factor to bring the mean free path below the injection length of turbulence. The higher the boosting factor is, the less the mean free path is obtained for the particles.

\begin{figure}[h]
\includegraphics[width=\columnwidth]{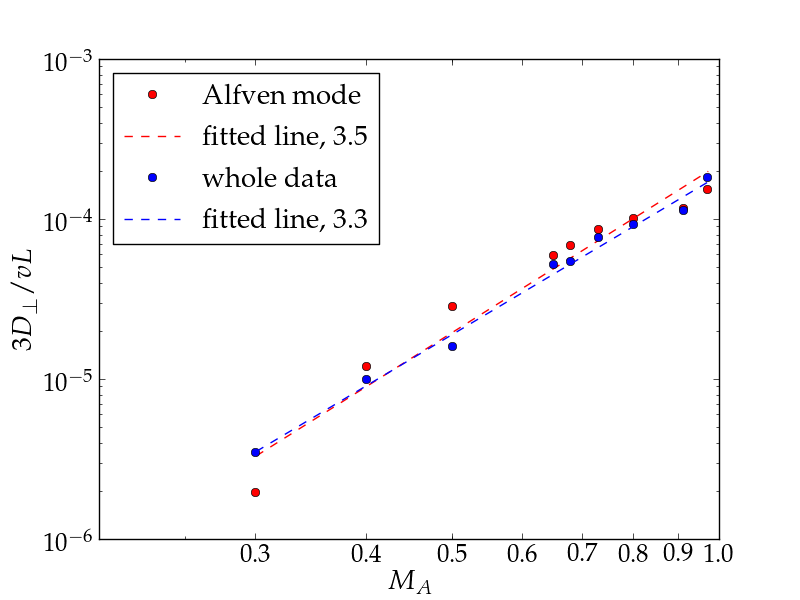}
%\label{Fig:dperpMa}
\centering
\caption{ The perpendicular diffusion coefficient with $\lambda_{\|} > L$ in MHD turbulence of various $M_A$. The results are obtained for the Alfv\'en and the whole turbulence data cubes.}\label{fig:dperp_mfpgtL}
\centering
\end{figure}

The dependence of the perpendicular diffusion coefficient on Mach number is presented in Fig.~\ref{fig:dperp_mfpgtL} for both total turbulence data cubes and the Alfv\'en modes. The relation between the diffusion coefficients and Alfv\'enic Mach numbers is fitted by a power law:$D_{\perp} \propto M_A ^\zeta$.
By taking into account the anisotropy of the Alfv\'enic turbulence, \citet{YL08} demonstrated that the relation between perpendicular diffusion $D_{\perp}$ and ${M_A}$ should have the power law index $\zeta=4$ instead of the $\zeta=2$ scaling calculated by \citet{KJ2000}. As is demonstrated by Fig.~\ref{fig:dperp_mfpgtL}, the index $\zeta$ is $3.5$ for Alfv\'en modes and $3.3$ for total turbulence data, respectively. Both results are more in favor of the YL08 calculation. 

\begin{figure}[h]
\includegraphics[width=\columnwidth]{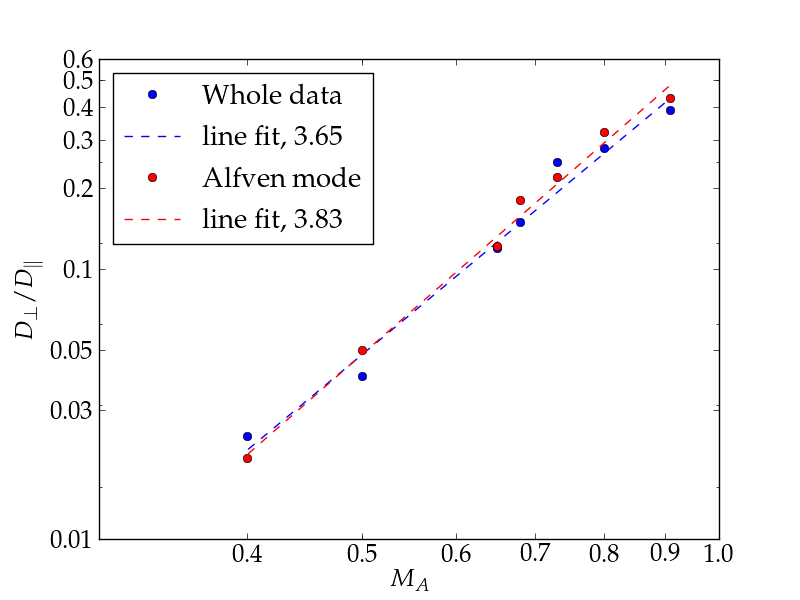}
\centering
\caption{The ratio between perpendicular and parallel diffusion ($D_\perp/D_\parallel$) when $\lambda_{\|} < L$. Transport of particles in both the whole turbulence (blue) and Alf\'ven modes (red) are presented. The fitting lines and power law indices are marked in the legend.
}\label{fig:dperpdpara_mfpltL}
\centering
\end{figure}

We further consider the CR propagation with mean free path smaller than the injection scale, which is the case for all Galactic CRs.
Simulations are executed with the artificial scattering included as aforementioned. In this regime, it is expected that the ratio between perpendicular and parallel diffusion coefficients ($D_\perp/D_\parallel$) will follow an $M_A^4$ dependence \citep{YL08}. The dependency of the diffusion coefficients on Mach numbers is presented in Fig.~\ref{fig:dperpdpara_mfpltL} for $\lambda_{\|} < L$. The diffusion coefficients are compared for both the Alfv\'en modes and the total turbulence. As demonstrated in Fig.~\ref{fig:dperpdpara_mfpltL}, the fitting index is $3.65$ for the total turbulence data cubes and $3.83$ for Alfv\'en modes. 

For both regimes where CRs' mean free path is larger and smaller than the injection scale, the results from Alfv\'en modes are closer to the expected index $\zeta=4$ (YL08) than those from total turbulence data cubes. This is due to the contributions from the magnetosonic modes in the total turbulence data cubes. Our calculations show that the CR perpendicular diffusion on large scale is strongly dependent on the Alfv\'enic Mach number, and it is essential to consider the anisotropy of MHD turbulence when modelling CR propagation.

\section{Perpendicular Transport on scales smaller than $L$ and super-diffusion} 

In this section, we will discuss the particle transport on small scales within the inertial range. The time evolution for the perpendicular transport can be fitted by a power law:

\begin{equation}\label{eq:dxvst}
d_\perp\equiv<(y-y_0)^2 +(z-z_0)^2>^{1/2}\propto t^{\alpha}
\end{equation}

The Richardson diffusion describes the explosive growth of the separation of particles in turbulence medium, as inferred from fluids experiments many decades ago \citep{Richardson26}. Richardson law is equivalent to the Kolmogorov spectrum. Therefore Richardson diffusion is also expected in MHD turbulence since the perpendicular spectrum of Alfv\'enic turbulence has a Kolmogorov scaling \citep{GS95}. The Richardson diffusion in MHD turbulence was confirmed with high resolution numerical simulations by \citet{Eyink_NAT}. Following the Richardson diffusion of magnetic field lines, CRs also undergo superdiffusion on the scales below the injection scale with the index over time $\alpha={3/2}$ \citep{YL08, LY14}.  

\begin{table}[ht]
\begin{center}
\begin{tabular}{ cccccccc  }
\hline
\hline
$M_A$ & 0.44 & 0.56 & 0.65 & 0.83 & 1.28 &1.4 &1.54 \\
\hline
 Global   & 0.95    & 1.1   &   1.2 & 1.15 & 1.25 & 1.3  & 1.6 \\
 Local    & 1.3     & 1.45  &   1.5 & 1.5  & 1.35 & 1.45 & 1.7 \\
\hline
\end{tabular} 
\end{center}
\caption{ 
The obtained super-diffusion index $\alpha$ is listed in the table. The diffusion coefficients are calculated and compared in both the global and local frames of reference. CR undergoes super-diffusion and the slope obtained is closer to 1.5 in the {\em local} reference frame.}\label{tab:index_cho}
\end{table}

\begin{figure}[h]
\includegraphics[width=\columnwidth]{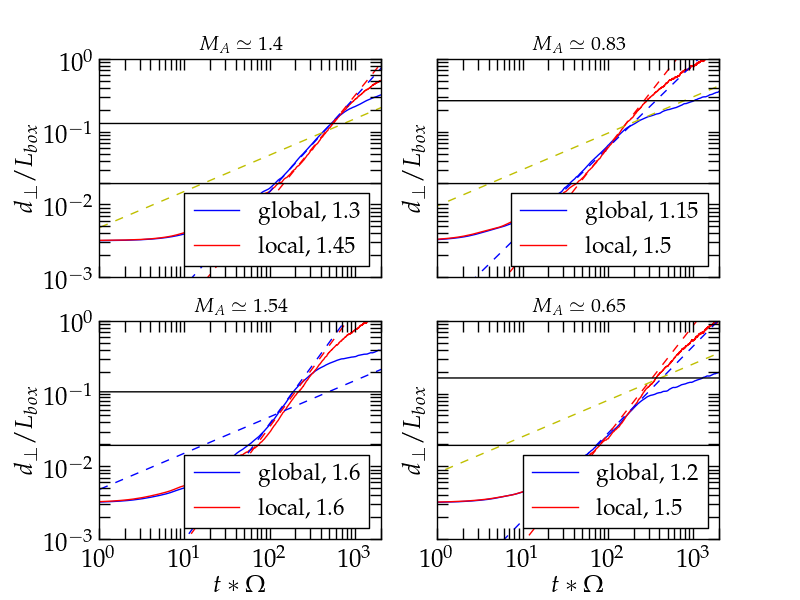}
\centering
\caption{ Perpendicular transport of CRs on small scales. The y axis represents the perpendicular distances normalised by the box length ($d_\perp/L_{box}$) and the x axis represents the CR gyro periods ($t*\Omega$). The perpendicular distances obtained from numerical simulations are represented in the global (blue lines) and the local (red lines) reference frame. The horizontal lines in the plots represents the inertial range of turbulence. The yellow lines represents the reference line for normal diffusion with a slope of 0.5.}\label{fig:dperp_cho}
\centering
\end{figure} 

%\begin{figure}[h]
%\includegraphics[width=0.49\columnwidth]{PSD_PENCIL_CRPAPER.png}
%\includegraphics[width=0.49\columnwidth]{curvefitF_PvsrA1.png}
%\centering
%\caption{\colz {\em Left)} The power spectrum density of turbulence %data {\em Right)}: The Probability distribution of test particles %launched from the same turbulence data cube vs. the distance r at %given time snapshots. The fitting is consistent with Kolmogorov %spectrum index h=1/3.}
%\label{PDF}
%\end{figure}

\begin{figure}[h]
\includegraphics[width=1.0\columnwidth, height=0.15\textheight]{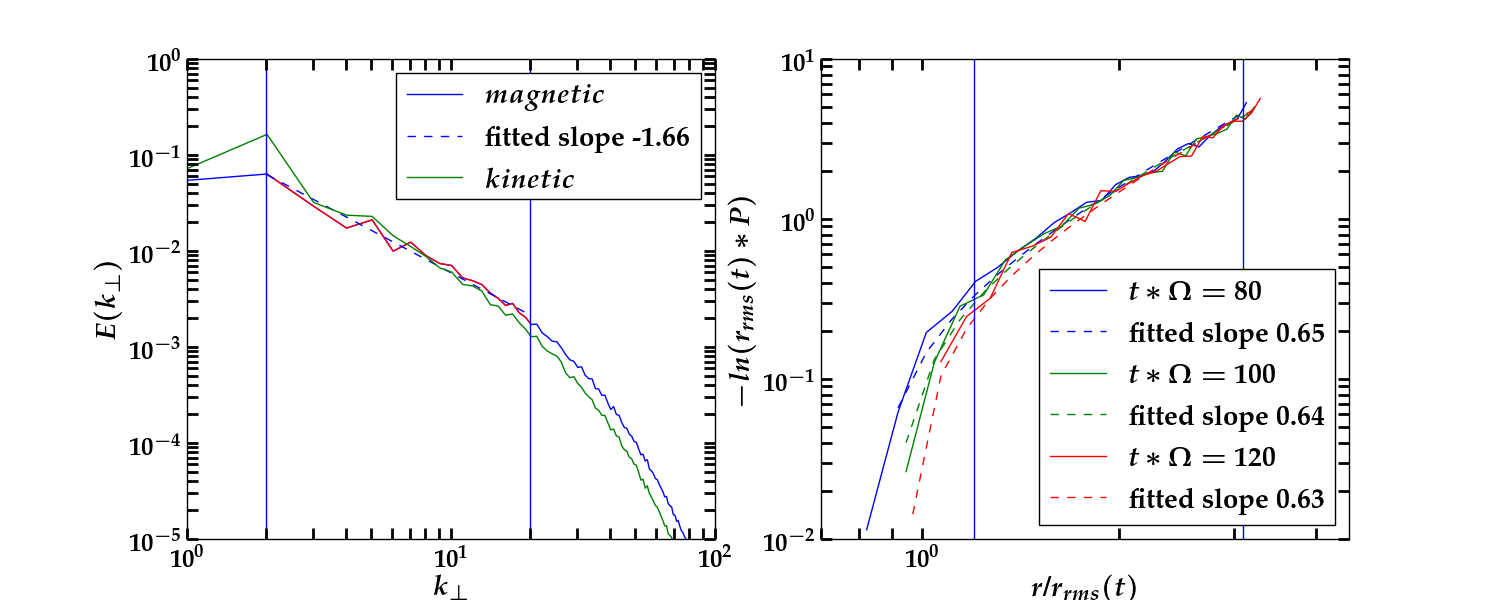}
\centering
\caption{\em Left) The power spectrum density of turbulence data {\em Right)}: The Probability distribution of test particles launched from the same turbulence data cube vs. the distance r at given time snapshots. The fitting is consistent with Kolmogorov spectrum index h=1/3.}
\label{PDF}
\end{figure}

%\begin{figure}[h]
%\includegraphics[width=0.49\columnwidth, %height=0.16\textheight]{MA073_Glob_all.png}
%\includegraphics[width=0.49\columnwidth,height=0.16\text%height]{MA073_Local_all.png}
%\centering
%\caption{The same as Fig.~\ref{fig:dperp_cho}, but with %$M_A\sim 0.73$ and results in decomposed modes included %for comparison.
%}\label{fig:dpara_totalAlf}
%\centering
%\end{figure}

\begin{figure}[h]
\centering
\includegraphics[width=1.0\columnwidth, height=0.15\textheight]{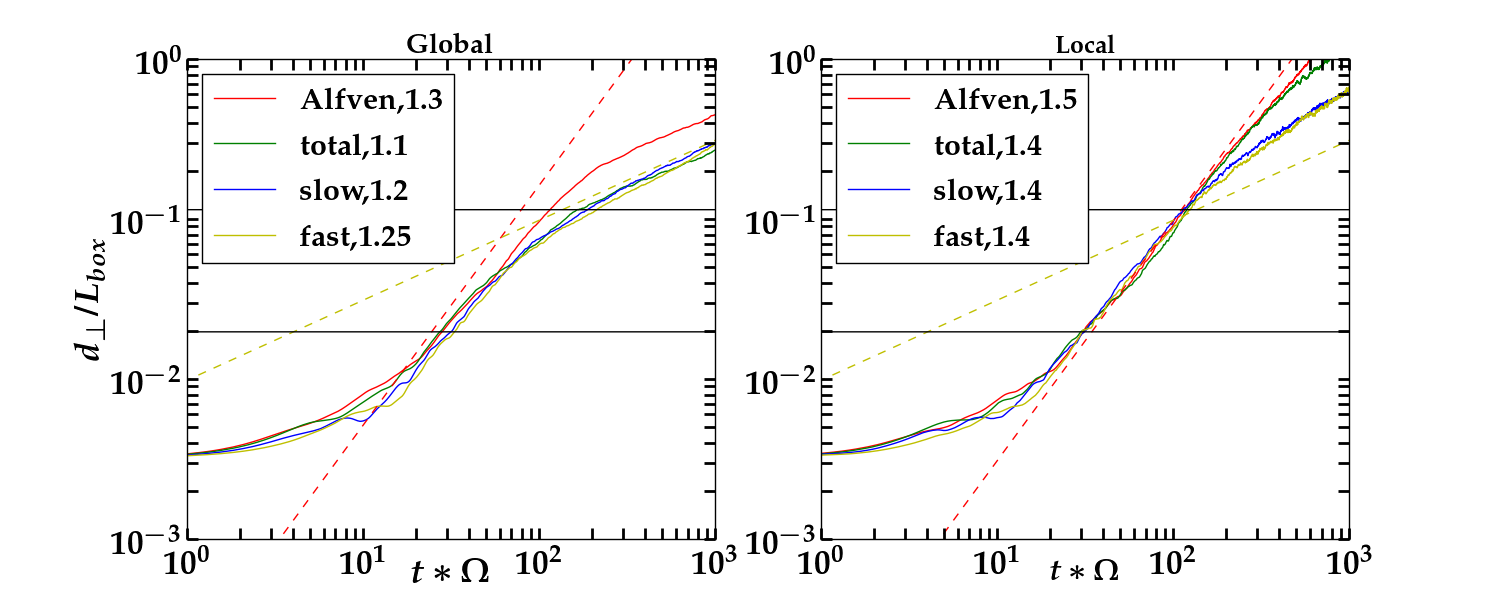}
\caption{The same as Fig.~\ref{fig:dperp_cho}, but with $M_A\sim 0.73$ and results in decomposed modes included for comparison.
}\label{fig:dpara_totalAlf}
\centering
\end{figure}

For the study of CR transport on small scales, test particle simulations are performed in MHD turbulence data cubes with CRs being initially grouped together into beams. The simulation box has $512^3$ cells. This box is divided into 64 equal sized cubes of size $128^3$ cells. From the centre of each of these cubes, a beam is launched parallel to the local magnetic field. The starting points of the beam particles are uniformly distributed around the centre of each cube with a separation of 1 cell unit among them. The CRs in each beam are very closely spaced and their initial pitch angles are set at zero so that it could be analyzed how the particle separation diverges with time. For a particular Mach number, the test particle simulations are done in different MHD modes as well as the total turbulence data cubes. There are two reference frames in the current simulations: the global reference frame, defined by the mean magnetic field of the turbulence data cube, and the local reference frame, defined by the mean magnetic field along the particle trajectory. The evolution of perpendicular CR transport in this work is calculated in both the global and the local frames of reference. The particle position, velocity and the magnetic field at the corresponding position are recorded at each time instance from test particle simulations. The values of the perpendicular distance is obtained by averaging over all combinations of pairs of particles within individual beams and then over all the beams.

The perpendicular distance $d_\perp$ is plotted against the propagation time of the CRs in Fig.~\ref{fig:dperp_cho} for sub-Alfv\'enic and super-Alfv\'enic total turbulence data. The results are fitted with Eq.~\eqref{eq:dxvst}.
Fig.~\ref{fig:dperp_cho} demonstrates the CR transport in the two reference frames: all cases can be fitted with the index close to $\alpha\sim1.5$ in the {\em local} reference frame; whereas for global reference frame, $\alpha$ is only close to 1.5 at super-Alfv\'enic cases, decreasing substantially and close to $1$ for sub-Alfv\'enic cases. In Table~\ref{tab:index_cho}, the comprehensive fitting indices are listed for the CR superdiffusion in total MHD turbulence data cubes with the Alfv\'enic Mach number ranging from 0.44 to 1.54. The fitting index is close to Richardson-diffusion $\alpha\sim1.5$ for all data measured in the {\em local} reference frame, which is in line with the theoretical expectations since the global magnetic field generally differs from local magnetic fields in turbulent medium.  Fig.~\ref{PDF} shows the power spectral density (PSD) of the turbulence data and the probability density functions (PDFs) of the test particles launched in the same turbulence data cube vs. distance r between them at different time snapshots. The distribution fits well to an exponential form, i.e., $P\propto \exp(-Cr^{1-h})$, the index $h=1/3$ is the Kolmogorov scaling, consistent with the Richardson diffusion \citep{Hentschel1984,Eyink_NAT}. 

We further study the perpendicular transport on small scales from decomposed MHD modes. As an example, Fig.~\ref{fig:dpara_totalAlf} demonstrates how we obtain the fitting index for the time evolution of the perpendicular transport in different MHD modes. Both local and global reference frames are used in our calculations. Turbulence data cubes with Alfv\'enic Mach numbers ranging from $0.4$ to $1.0$ are considered and the fitting power law index $\alpha$ are presented for total and decomposed modes in Table~\ref{tab:index_modecompare_pencilsubAlf}. Superdiffusion is generally observed in all our tests \footnote{We randomly selected a few cases to repeat the test particle simulations in a higher resolution turbulence data ($1024^3$) with the same $M_A$. The results do not show obvious difference.}. From Table~\ref{tab:index_modecompare_pencilsubAlf}, we find that the particles in decomposed Alfv\'enic modes are the closest to the Richardson diffusion (index $\alpha=1.5$) in the {\em local} reference frame compared to the other modes. In global magnetic reference frame, the indices deviate further from the Richardson diffusion as expected. It implies that the observed superdiffusion index can vary determined by the modes composition as well as the Alfv\'enic Mach number of the local turbulence.
 
\begin{table}[h]
\begin{center}
\begin{tabular}{ccccccccc}
\hline
\hline
 $M_A$ & 0.4 & 0.5 & 0.65 & 0.68 & 0.73  & 0.8 & 0.91  & 0.97 \\
\hline
Global MM   & 0.8   & 1.0    &  1.1  & 1.1  & 1.1  & 1.2  & 1.15 & 1.15  \\
Global AM   & 1.1   & 1.65   &  1.3  & 1.3  & 1.3  & 1.35 & 1.3  & 1.4   \\
Global SM  & 0.95  & 1.3    &  1.3  & 1.1  & 1.2  & 1.3  & 1.2  & 1.3   \\
Global FM  & 1.1   & 1.2    &  1.1  & 0.9  & 1.25 & 1.25 & 1.6  &1.4    \\
Local  MM   & 1.2   & 1.5    &  1.6  & 1.4  & 1.4  & 1.5  & 1.4  &1.4    \\
Local  AM  & {\bf 1.5}   & {\bf 1.5}    &  {\bf 1.5}  & {\bf 1.5}  & {\bf 1.5} & {\bf 1.5}  & {\bf 1.5}  & {\bf 1.6}    \\
Local  SM & 1.6   & 1.4    &  1.3  & 1.3  & 1.4  & 1.4  & 1.3  & 1.3   \\
Local  FM & 1.55  & 1.7    &  1.1  & 1.3  & 1.4  & 1.4  & 1.7  & 1.4   \\
\hline    
\end{tabular}
\end{center}
\caption{The same as Table~\ref{tab:index_cho} but the comparison extends to decomposed MHD modes. AM represents Alfv\'en mode, SM represents slow modes, FM represents fast mode and MM represents the total data cube. ``Global'' and ``Local'' represent the magnetic reference frame in the calculation.}\label{tab:index_modecompare_pencilsubAlf}
\end{table}

We also calculate the $M_A$ dependence of the super-diffusion ${d_{\perp}}^2 / t^3 $ of particles. It is expected that for sub-Alfv\'enic turbulence the dependence is $M_A^4$ and for super-Alfv\'enic turbulence the dependence is $M_A^3$ \citep{YL08,LY14}. 

\begin{figure}[h]
\includegraphics[width=9 cm]{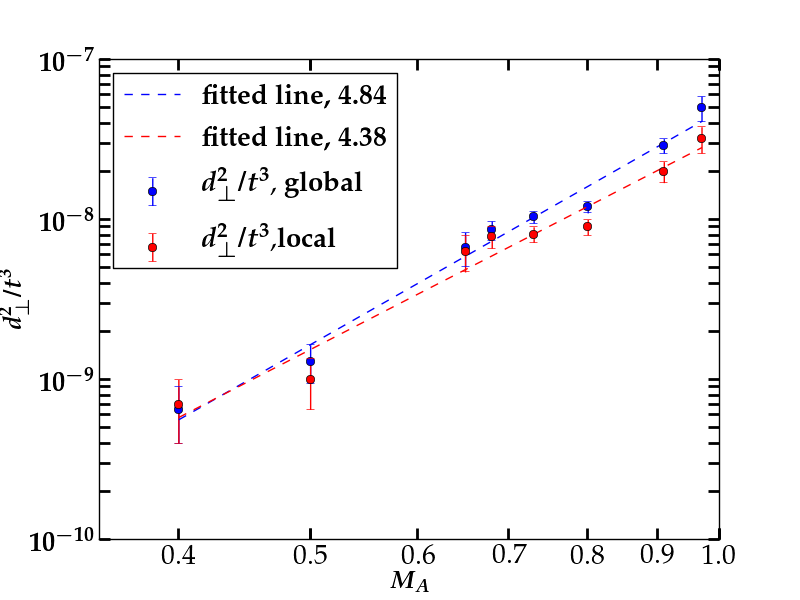}
\centering
\caption{Superdiffusion of particles with $\lambda_{\|} > L$ in Alfv\'en modes. The blue line is the fit in the global reference frame. The red line shows the fit for the data points obtained in the local reference frame.}\label{fig:diffindexvsMA_subAlf}
\centering
\end{figure}

Fig.~\ref{fig:diffindexvsMA_subAlf} demonstrates the CR perpendicular diffusion in Alfv\'en modes in sub-Alfv\'enic regime. The diffusion coefficients are calculated in both the local and global reference frames. The fitting power law index in the {\em local} reference frame ($4.34$) is closer to the theoretical expectation $ M_A ^4$ than that in the global frame ($4.84$). 

\begin{figure}[h]
\includegraphics[height=6cm, width=\columnwidth]{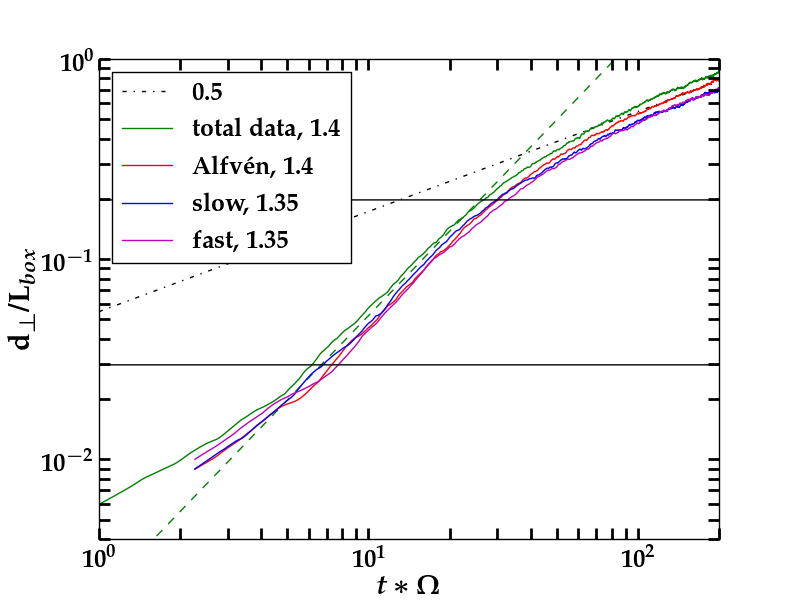}
\centering
\caption{The same as Fig.~\ref{fig:dpara_totalAlf}, but in local reference frame with super-Alfv\'enic turbulence $M_A=2.11$.
}\label{fig:superdiff_superAlf}
\centering
\end{figure}

\begin{figure}[h]
\includegraphics[width=\columnwidth]{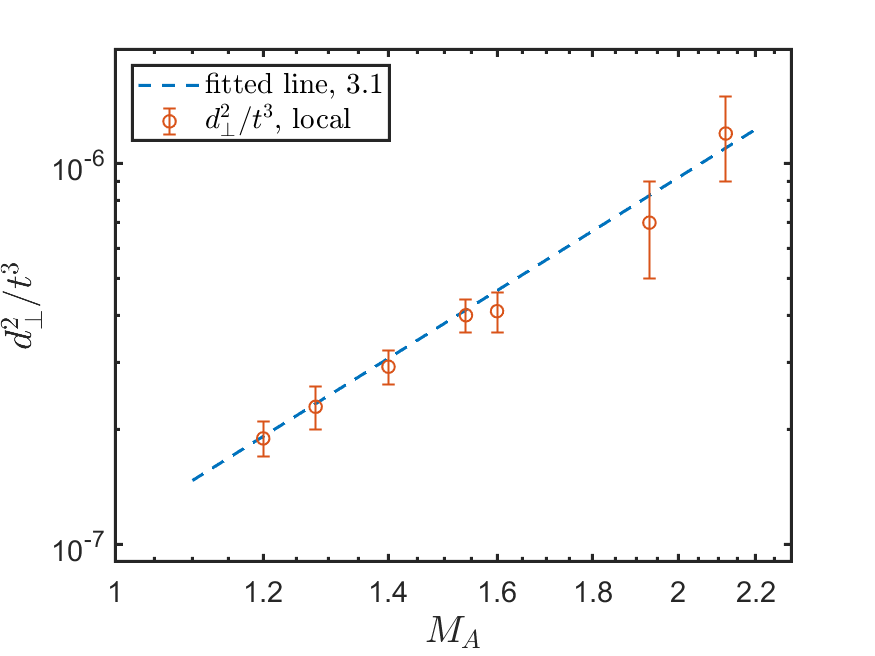}
\centering
\caption{The same as Fig.~\ref{fig:diffindexvsMA_subAlf}, but with super-Alfv\'enic turbulence. The calculation is done in local reference frame.
}\label{fig:diffindexvsMA_superAlf}
\centering
\end{figure}

\begin{figure}
    \centering
    \includegraphics[width=\columnwidth]{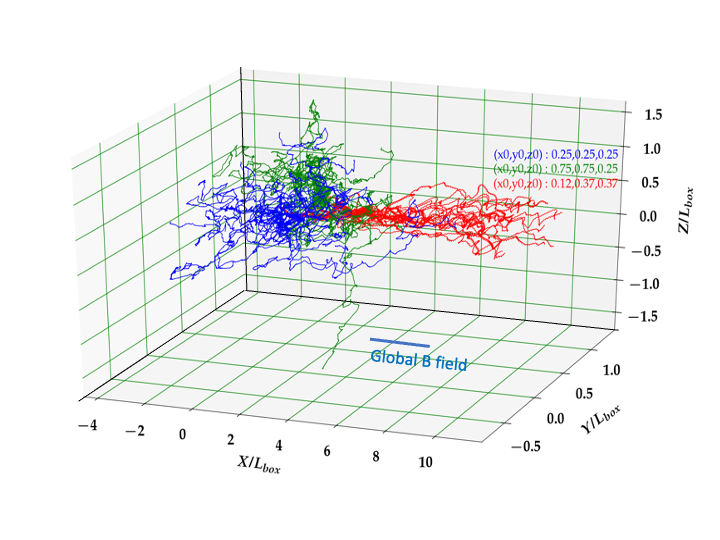}
    \caption{3D trajectories of particles launched from a beam. They experience super-diffusion because of Richardson diffusion of magnetic fields in turbulence. The rate of superdiffusivity, the $\alpha$ index, depends on the ratio of mean free path to the injection scale. See also Fig.\ref{fig:superdiff_subAlf}.}
    \label{fig:3D_traj}
\end{figure}

\begin{figure}[h]
\includegraphics[width=\columnwidth]{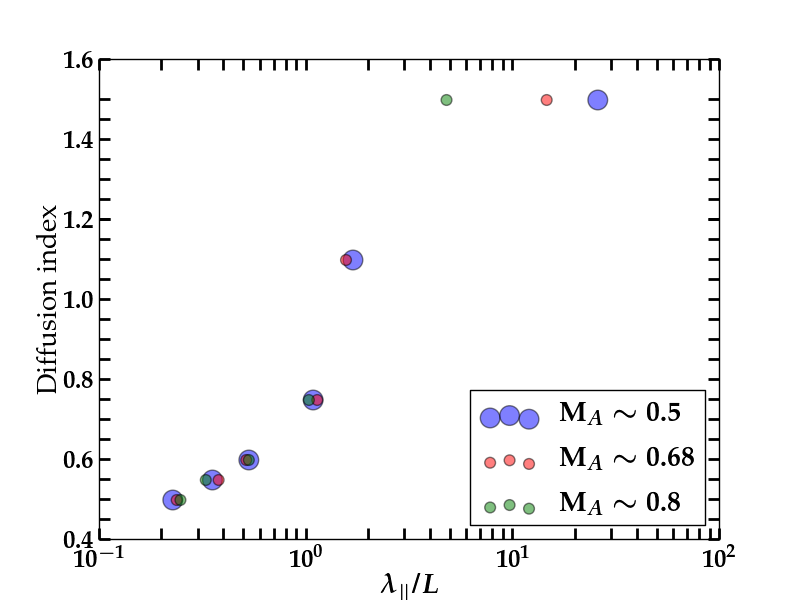}
\caption{The diffusion index $\alpha$ vs. $\lambda_{\|} $ in the {\em local} reference frame.}\label{fig:superdiff_subAlf}
\centering
\end{figure}

Fig.~\ref{fig:superdiff_superAlf} shows the perpendicular diffusion in super-Alfv\'enic turbulence with total data cube and decomposed MHD modes. The inertial range of MHD turbulence starts from $L M_A^{-3}$. The MHD mode decomposition is only performed within the inertial range. 
We fit the time evolution of perpendicular transport with Eq~\eqref{eq:dxvst} when the particles are within the inertial range (indicated by the horizontal lines). The CRs in the decomposed Alfv\'en modes have shown the super-diffusion with power law index $1.5$. The contributions from magnetosonic modes result in the slightly different power law index obtained in the total turbulence data ($1.4$).

Fig.\ref{fig:diffindexvsMA_superAlf} represents $M_A$ dependence of the perpendicular diffusion at the {\em local} reference frame in different super-Alfv\'enic turbulence. The super-diffusion on small scales ${d_{\perp}}^2 / t^3 $ in the {\em local} reference frame show a dependence of $M_A^{3.1}$ in our calculations, close to the $ M_A^3$ theoretical relation.

 We also studied the dependence of the superdiffusion on the mean free path $\lambda_\|$ of the particles. The 3D trajectories of the particles in the MHD turbulence data are presented in Fig.~\ref{fig:3D_traj}. Three beams are launched from three randomly selected positions. The (red) beam with larger mean free path ($\lambda_\|\simeq 15L$) exhibits superdiffusion guided by magnetic field lines. In contrast to hydrodynamic system, the spread is substantially smaller, consistent with earlier study \citep{Eyink_NAT}. On the other hand, the (green) beam launched with mean free path ($\lambda_\|\simeq 0.9L$) displays more stochasticity, in line with diffusion/random walk process. The blue beam has an intermediate mean free path ($\lambda_\|\simeq 1.5L$) and shows a characteristics in between. Fig.~\ref{fig:superdiff_subAlf} displays the diffusion index $\alpha$ of CRs as a function of mean free path $\lambda_{\|}$ in the {\em local} reference frame.  As we see there, the index is dictated by the ratio of $\lambda_{\|}$ to the injection scale L. It changes smoothly from 0.5 corresponding to the normal diffusion regime in the case of $\lambda_{\|}/L\ll 1$ to 1.5 when $\lambda_{\|}/L\gg 1$ in correspondence to Richardson diffusion. Particularly the super-diffusion becomes $0.75$ when $\lambda_\|\sim L$. 

\section{Discussion}

Cosmic ray transport is intimately linked to the property of MHD turbulence. Different from hydrodynamic turbulence, MHD turbulence is much widely diversified depending on the parameters in local interstellar environment, such as Mach number and plasma $\beta$. Another factor, that has been frequently overlooked, is the modes composition of MHD turbulence. It can vary substantially depending on the driving mechanism of turbulence \citep{KY20}. This is particularly important in view of the fact that different MHD modes contribute to CR transport differently. It is therefore inadequate to depict CR transport as that described by Kolmogorov turbulence with one characterization even in the high energy regime where external turbulence dominates the CR scattering.

The pitch angle scattering and therefore parallel diffusion is dominated by fast modes, as confirmed by our test particle simulations here. The parallel diffusion coefficient varies, thereby, with the percentage of fast modes and the forcing mechanism of the local turbulence \citep{Zhang20NA}. On the other hand, the cross field transport is much determined by Alfv\'en modes. As demonstrated in the paper, the tests performed with Alfv\'en modes show better consistency with theoretical predictions earlier \citep{YL08, LY14} particularly in the {\em local} reference frame. Then depending on the degree of Alfv\'enicity (the proportion of Alfv\'en modes in local MHD turbulence), the observed cross field transport property can vary.  The super-diffusion index can deviate from the Richardson diffusion determined by the Alfv\'enicity and the mean free path of the particles. This explains the observed diversity of super-diffusion indices \citep[see, e.g.][]{Perri2016}. The specific $M_A$ dependence can also show some deviation from the theoretical values, e.g., $M_A^4$ in the case of sub-Alfv\'enic turbulence and $M_A^3$ for the super-Alfv\'enic turbulence.

Damping also plays an important role in shaping the CR diffusion properties, especially the energy dependence. It is, nonetheless, not covered in the test particle simulations with MHD turbulence since damping physics, particularly, the collisionless damping can not be captured in MHD. We, therefore, do not pay particular attention to the energy dependence of the transport properties. It will be subject for future studies.

\section{Summary}

In this paper, we have carried out test particle simulations to study the diffusion of CRs in different MHD turbulence. The particles are considered with the mean free path $\lambda_{\|}$ both larger and smaller than the injection scale $L$. The MHD turbulence data cubes range from sub-Alfv\'enic to super-Alfv\'enic regimes. The test particle simulations are also performed in the three MHD modes (Alfv\'en, slow and fast), decomposed from the MHD turbulence data cubes. The CRs propagating within and beyond the inertial range are investigated. The test particle results are examined in both the local and global reference frames. Our main results are summarised below.

\begin{enumerate}

\item The pitch angle scattering test of CRs for different MHD modes shows that: while the pitch angle scattering in Alfv\'en and slow modes show similar diffusion coefficients, the fast modes differ and are much more efficient in CR scattering. 

\item Cross field transport of particles is normal diffusion on large scales. The ratio between perpendicular and parallel diffusion coefficients is close to the $M_A^4$ dependence.

\item Particles undergo superdiffusion on scales smaller than injection scale of turbulence. The super-diffusion rate $d_\perp^2/t^3$ has shown a strong dependence on the  Alfv\'enic Mach number.

\item Richardson super-diffusion is well recovered (1.5 for for $\lambda_{\|}>L$ and  reduces with the decrease of $\lambda_{\|}$. In particular, the super-diffusion index becomes 0.75 for $\lambda_{\|}\sim L$ and normal diffusion is recovered when $\lambda_{\|}\ll L$ in the {\em local} magnetic reference frame with Alfv\'en modes decomposed from simulated turbulence data. The actual observed super-diffusion index also varies with the modes composition.
\end{enumerate}

\acknowledgements
The authors thank the anonymous referee for their valuable comments, which has helped improve the quality of the paper. We acknowledge helpful communications with S. Malik.

\bibliographystyle{aasjournal}
\bibliography{CRpaper}
\appendix
\setcounter{figure}{0}
\renewcommand{\thefigure}{A\arabic{figure}}

\section{Diffusion coefficient for CR transport on global scales}

\begin{figure}[h]
\centering
\includegraphics[width=0.45\columnwidth, height=0.28\textheight]{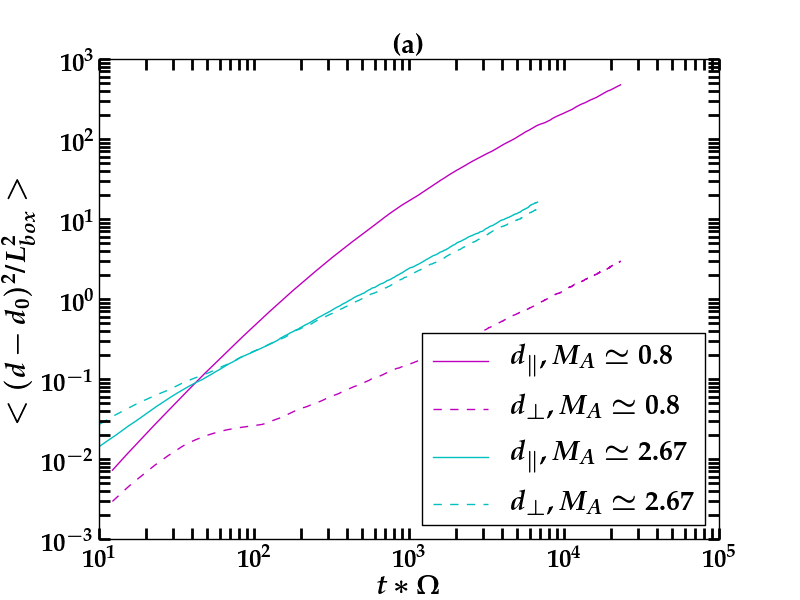}
\includegraphics[width=0.45\columnwidth, height=0.28\textheight]{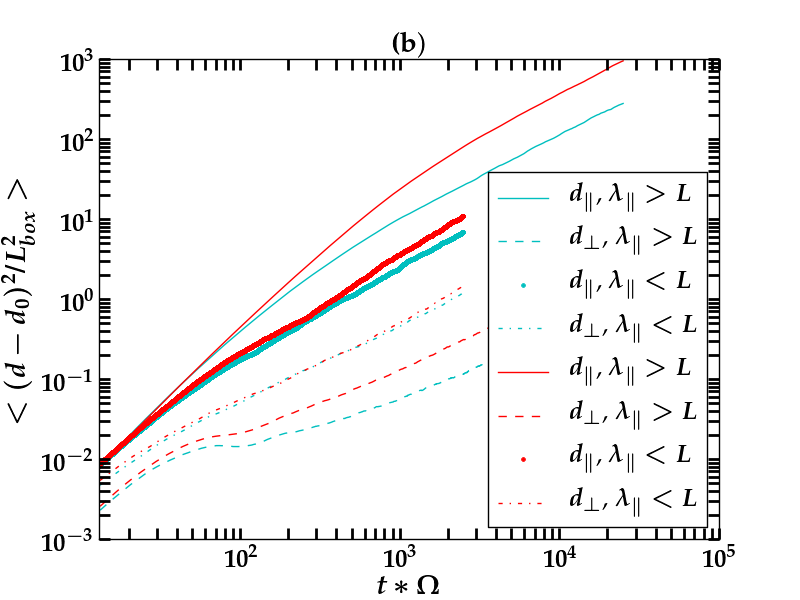}
\caption{(a): Mean square distance traveled by the particles vs. time. Both sub-Alfv\'enic ($M_A\simeq0.8$, magenta) and super-Alfv\'enic ($M_A\simeq$2.67, cyan) turbulence are presented. The y axis represents the mean square distance $<(d-d_0)>^2$ of CR transport normalised by the box length, $L_{box}^2$. The particle running is in the $\lambda_\|>L$ regime. The x axis represents the gyro periods $t*\Omega$. The parallel and perpendicular distances obtained from numerical simulations are represented by bold and dashed line, respectively. The particles have shown normal-diffusion. The diffusion becomes isotropic in the super-Alfv\'enic turbulence; (b): Same for both CRs with $\lambda_{\|} > L$ and with $\lambda_{\|} < L$ in total turbulence data (green) and Alfv\'en modes (red) with $M_A\simeq$0.73. After introducing artificial scattering, the transport becomes more isotropic compared to no artificial scattering.}\label{fig:dxx}
\end{figure}

In this section, we present our calculations for the diffusion of CRs on large scales beyond the inertial range. Fig.~\ref{fig:dxx}a demonstrates the parallel and perpendicular distance for CR propagation in sub-/super-Alfv\'enic turbulence. Fig.~\ref{fig:dxx}b compares the ensemble-averaged square distance of the particles when the mean free path is greater or smaller than the injection scale. In this regime, all our simulations have made sure that the running time is sufficient and normal diffusion is observed. To perform parallel and perpendicular diffusion coefficients, we take the range when the linear growth is observed in the time evolution figure. We note that in Fig.~\ref{fig:dxx}a, the parallel diffusion is much larger than the perpendicular diffusion for $M_A\simeq 0.8$ but they are equal to each other for $M_A\simeq 2.67$. This difference demonstrates the anisotropy is presented in sub-Alfv\'enic turbulence but the super-Alfv\'enic turbulence is almost isotropic.

\end{document}